\documentclass[titlepage,12pt]{article}
\usepackage{amssymb}
\usepackage{amsfonts}
\usepackage{geometry}
\usepackage{mathrsfs}
\usepackage[onehalfspacing]{setspace}

\input{tcilatex}
\renewcommand{\mathcal}{\mathscr}

\begin{document}

\title{Multiplicative component models for replicated point processes}
\author{Daniel Gervini \\
Department of Mathematical Sciences\\
University of Wisconsin--Milwaukee}
\maketitle

\begin{abstract}
We propose a multiplicative semiparametric model for the intensity function
of replicated point processes. Two examples of applications are given: a
temporal one, about the dynamics of Internet auctions, and a spatial one,
about the spatial distribution of street robberies in Chicago.

\end{abstract}

\section{Introduction}

Point processes in time and space have a broad range of applications, in
diverse areas such as neuroscience, ecology, finance, astronomy, seismology,
and many others. Examples are given in classic textbooks like Cox and Isham
(1980), Diggle (2013), M\o ller and Waagepetersen (2004), Streit (2010), and
Snyder and Miller (1991), and in the papers cited below. However, the
point-process literature has mostly focused on single-realization cases,
such as the distribution of trees in a single forest (Jalilian et al., 2013)
or the distribution of cells in a single tissue sample (Diggle et al.,
2006). Situations where several replications of a process are available are
increasingly common, but this area is still relatively unexplored in the
literature. We can cite Diggle et al.~(1991), Baddeley et al.~(1993), Diggle
et al.~(2000), Bell and Grunwald (2004), Landau et al.~(2004), Wager et
al.~(2004), and Pawlas (2011). However, these papers propose estimators for
summary statistics of the processes rather than the intensity functions,
which would be more informative.

When several replications of a process are available, it is possible to
estimate the intensity functions by \textquotedblleft borrowing
strength\textquotedblright\ across replications. Along these lines Wu et
al.~(2013) propose estimators for the mean and principal components of
independent and identically distributed realizations of a temporal doubly
stochastic process based on kernel estimators of covariance functions.
Gervini (2016) proposes an additive independent component model that has the
advantages, over Wu et al., of treating the temporal and spatial cases in a
unified way and of being easy to extend beyond the i.i.d.~case, for
instance, to regression and multivariate settings. In fact, Gervini and Baur
(2017) is an extension of this method to marked point processes.

In this paper we propose an alternative to the additive model of Gervini
(2016), namely an additive model for the log-intensity functions. This
simplifies the numerical and theoretical aspects of the procedure by
eliminating the nonnegativity constraints, but the interpretability is
somewhat hampered by the fact that the additive model for the
log-intensities translates into a multiplicative model for the intensities.
At the end of this brief paper we present two examples of application, one
temporal and one spatial, to illustrate these issues.

\section{The model\label{sec:Model}}

A point process $X$ is a random countable set in a space $\mathcal{S}$,
where $\mathcal{S}$ is usually $\mathbb{R}$ for temporal processes and $%
\mathbb{R}^{2}$ or $\mathbb{R}^{3}$ for spatial processes (M\o ller and
Waagepetersen, 2004, ch.~2; Streit, 2010, ch.~2). A process is locally
finite if $\#(X\cap B)<\infty $ with probability one for any bounded $%
B\subseteq \mathcal{S}$. In that case we can define the count function $%
N(B)=\#(X\cap B)$ for any bounded $B\subseteq \mathcal{S}$, which
essentially characterizes the process and is equivalent to $X$ in this case.

Let $X$ be locally finite and define $X_{B}=X\cap B$. Given a locally
integrable function $\lambda :\mathcal{S}\rightarrow \lbrack 0,\infty )$,
i.e.~a function $\lambda $ such that $\int_{B}\lambda <\infty $ for any
bounded $B\subseteq \mathcal{S}$, we say that $X$ is a Poisson process with
intensity function $\lambda $, denoted by $X\sim \mathcal{P}(\lambda )$, if 
\emph{(i)} $N(B)$ follows a Poisson distribution with rate $\int_{B}\lambda $
and \emph{(ii)} conditionally on $N(B)=m$, the $m$ points in $X_{B}$ are
independent and identically distributed with density $\tilde{\lambda}%
=\lambda /\int_{B}\lambda $.

For $X\sim \mathcal{P}(\lambda )$, then, the density function of $X_{B}$ at $%
x_{B}=\{t_{1},\ldots ,t_{m}\}$ is 
\begin{eqnarray}
f(x_{B}) &=&f(m)f(t_{1},\ldots ,t_{m}|m)  \label{eq:Pois_lik} \\
&=&\exp \left\{ -\int_{B}\lambda (t)dt\right\} \frac{\{\int_{B}\lambda
(t)dt\}^{m}}{m!}\times \prod_{j=1}^{m}\tilde{\lambda}(t_{j})  \nonumber \\
&=&\exp \left\{ -\int_{B}\lambda (t)dt\right\} \frac{1}{m!}%
\prod_{j=1}^{m}\lambda (t_{j}).  \nonumber
\end{eqnarray}%
What we mean by\ density of $X_{B}$, whose realizations are sets, not
vectors, is the following: if $\mathcal{N}$ is the family of locally finite
subsets of $\mathcal{S}$, i.e.~$\mathcal{N}=\{A\subseteq \mathcal{S}%
:\#(A\cap B)<\infty $ for all bounded $B\subseteq \mathcal{S}\}$, then for
any $F\subseteq \mathcal{N}$, 
\begin{eqnarray*}
P\left( X_{B}\in F\right) &=&\sum_{m=0}^{\infty }\int_{B}\cdots \int_{B}%
\mathbb{I}(\{t_{1},\ldots ,t_{m}\}\in F)f(\{t_{1},\ldots
,t_{m}\})dt_{1}\cdots dt_{m} \\
&=&\sum_{m=0}^{\infty }\frac{\exp \left\{ -\int_{B}\lambda (t)dt\right\} }{m!%
}\int_{B}\cdots \int_{B}\mathbb{I}(\{t_{1},\ldots ,t_{m}\}\in
F)\{\prod_{j=1}^{m}\lambda (t_{j})\}dt_{1}\cdots dt_{m},
\end{eqnarray*}%
and, more generally, for any function $h:\mathcal{N}\rightarrow \lbrack
0,\infty )$ 
\begin{equation}
E\{h(X_{B})\}=\sum_{m=0}^{\infty }\int_{B}\cdots \int_{B}h(\{t_{1},\ldots
,t_{m}\})f(\{t_{1},\ldots ,t_{m}\})dt_{1}\cdots dt_{m}.  \label{eq:Eh}
\end{equation}%
A function $h$ on $\mathcal{N}$ is a function well defined on $\mathcal{S}%
^{m}$ for any integer $m$ and invariant under permutation of the
coordinates; for example, $h(\{t_{1},\ldots ,t_{m}\})=\sum_{j=1}^{m}t_{j}/m$.

Single realizations of point processes are often modeled as Poisson
processes with fixed $\lambda $s, but for replicated point processes a
single intensity function $\lambda $ rarely provides an adequate fit for all
replications. It is more reasonable to assume that the $\lambda $s are
subject-specific and treat them as latent random effects. Such processes are
called doubly stochastic or Cox processes (M\o ller and Waagepetersen, 2004,
ch.~5; Streit, 2010, ch.~8). A doubly stochastic process is a pair $%
(X,\Lambda )$ where $X|\Lambda =\lambda \sim \mathcal{P}(\lambda )$ and $%
\Lambda $ is a random function that takes values on the space $\mathcal{F}$
of non-negative locally integrable functions on $\mathcal{S}$. The $n$
replications of the process are then i.i.d.~realizations $(X_{1},\Lambda
_{1}),\ldots ,(X_{n},\Lambda _{n})$ of $(X,\Lambda )$, where $X$ is
observable but $\Lambda $ is not. In this paper we will assume that all $%
X_{i}$s are observed on a common region $B$ of $\mathcal{S}$; the method can
be extended to $X_{i}$s observed on non-conformal regions $B_{i}$ at the
expense of higher computational complexity.

The latent intensity process $\Lambda $ characterizes the distribution of $X$%
. Gervini (2016) proposes an additive model for $\Lambda $, but here we will
explore the alternative approach of assuming an additive model for $\log
\Lambda $, which is not constrained to be nonnegative. Let us assume, then,
that 
\begin{equation}
\log \Lambda (t)=\mu (t)+\sum_{k=1}^{p}U_{k}\phi _{k}(t)
\label{eq:Log_lambda_model}
\end{equation}%
where $\mu \in L^{2}(B)$ and $\phi _{1},\ldots ,\phi _{p}$ are orthonormal
functions in $L^{2}(B)$. The $U_{k}$s are assumed independent $N(0,\sigma
_{k}^{2})$ random variables. Model (\ref{eq:Log_lambda_model}), minus the
Gaussianity assumption, is a truncated version of the Karhunen--Lo\`{e}ve
expansion (Ash and Gardner, 1975, ch.~1) that any process in $L^{2}(B)$ must
follow, so it requires little justification. The Gaussianity assumption on
the $U_{k}$s is added in order to derive maximum likelihood estimators; see
next section. Model (\ref{eq:Log_lambda_model}) translates into a
multiplicative model for $\Lambda (t)$: 
\begin{equation}
\Lambda (t)=\lambda _{0}(t)\prod_{k=1}^{p}\xi _{k}(t)^{U_{k}},
\label{eq:Lambda_model}
\end{equation}%
where $\lambda _{0}=\exp \mu $ is the baseline intensity function and $\xi
_{k}=\exp \phi _{k}$ is a multiplicative component.

The mean and components of model (\ref{eq:Log_lambda_model}) are functional
parameters that need to be estimated. We will follow a semiparametric
approach, modeling $\mu $ and the $\phi _{k}$s in terms basis functions $%
\beta _{1},\ldots ,\beta _{q}$ which can be, for example, B-splines for
temporal processes or radial Gaussian kernels for spatial processes.
Simplicial bases are another possibility for spatial processes, particularly
if the domain $B$ is irregular. In any case, we will have $\mu (t)=\mathbf{c}%
_{0}^{T}\mathbf{\beta }(t)$ and $\phi _{k}(t)=\mathbf{c}_{k}^{T}\mathbf{%
\beta }(t)$, where $\mathbf{\beta }$ is the vector of the $\beta _{k}$s.
From (\ref{eq:Log_lambda_model}) we can express 
\[
\log \Lambda (t)=(\mathbf{c}_{0}+\mathbf{CU})^{T}\mathbf{\beta }(t) 
\]%
where $\mathbf{C}=[\mathbf{c}_{1},\ldots ,\mathbf{c}_{p}]$ and $\mathbf{U}%
=(U_{1},\ldots ,U_{p})^{T}$. The parameters $\mathbf{c}_{0}$ and $\mathbf{c}%
_{k}$s, along with the variances $\sigma _{k}^{2}$s of the $U_{k}$s, are
estimated by penalized maximum likelihood, as explained next.

\section{Estimation\label{sec:Estimation}}

Let us collect the parameters $\mathbf{c}_{0}$, $\mathbf{c}_{k}$s and $%
\sigma _{k}^{2}$s into a single vector $\mathbf{\theta }$. From now on we
will omit the subindex $B$ in $x_{B}$, since $B$ is fixed. Then the marginal
density of $X_{B}$ at $x$ is 
\begin{eqnarray}
f(x;\mathbf{\theta }) &=&\int \int f(x,\mathbf{u})~d\mathbf{u}
\label{eq:marg_XB} \\
&=&\int \int f(x\mid \mathbf{u})f(\mathbf{u})~d\mathbf{u}  \nonumber
\end{eqnarray}%
where, for $x=\{t_{1},\ldots ,t_{m}\}$, 
\begin{eqnarray*}
\log f(x\mid \mathbf{u}) &=&-\int_{B}\lambda _{\mathbf{u}}(t)dt+%
\sum_{j=1}^{m}\log \lambda _{\mathbf{u}}(t_{j})-\log m! \\
&=&-\int_{B}\exp \{(\mathbf{c}_{0}+\mathbf{Cu})^{T}\mathbf{\beta }(t)\}dt \\
&&+(\mathbf{c}_{0}+\mathbf{Cu})^{T}\sum_{j=1}^{m}\mathbf{\beta }(t_{j})-\log
m!
\end{eqnarray*}%
and 
\[
\log f(\mathbf{u})=\sum_{k=1}^{p}\left( -\frac{1}{2}\log 2\pi \sigma
_{k}^{2}-\frac{u_{k}^{2}}{2\sigma _{k}^{2}}\right) . 
\]%
There is no closed form for $f(x;\mathbf{\theta })$ but it can be easily
computed by Monte Carlo integration, as explained in the Technical
Supplement.

The model parameters are estimated by penalized maximum likelihood. Since
the dimension $q$ of the functional basis $\mathbf{\beta }$ may be large, a
roughness penalty is necessary to obtain smooth $\mu $ and $\phi _{k}$s. We
use penalties of the form $P(g)=\int_{B}\left\Vert \mathrm{H}g(t)\right\Vert
_{F}^{2}\ dt$, where $\mathrm{H}$ denotes the Hessian and $\left\Vert \cdot
\right\Vert _{F}$ the Frobenius matrix norm. Then for a temporal process $%
P(g)=\int (g^{\prime \prime })^{2}$ and for a spatial process $P(g)=\int \{(%
\frac{\partial ^{2}g}{\partial t_{1}^{2}})^{2}+2(\frac{\partial ^{2}g}{%
\partial t_{1}\partial t_{2}})^{2}+(\frac{\partial ^{2}g}{\partial t_{2}^{2}}%
)^{2}\}$, both of which are quadratic in the basis coefficients when
evaluated at $\mu $ and the $\phi _{k}$s.

Then the penalized maximum likelihood estimator $\mathbf{\hat{\theta}}$
based on $n$ independent realizations $x_{1},\ldots ,x_{n}$ is 
\[
\mathbf{\hat{\theta}}=\limfunc{argmax}_{\mathbf{\theta }}\rho _{n}(\mathbf{%
\theta }) 
\]%
where 
\[
\rho _{n}(\mathbf{\theta })=\frac{1}{n}\sum_{i=1}^{n}\log f(x_{i};\mathbf{%
\theta })-\nu _{1}P(\mu )-\nu _{2}\sum_{k=1}^{p}P(\phi _{k}) 
\]%
and $\nu _{1}$ and $\nu _{2}$ are smoothing parameters. We use two different
parameters for $\mu $ and the $\phi _{k}$s because the latter have unit norm
but $\mu $ does not, so it may be necessary to use $\nu _{1}$ and $\nu _{2}$
of different magnitudes to attain the same degree of smoothness. As
mentioned before, $P(\mu )=\mathbf{c}_{0}^{T}\mathbf{\Omega c}_{0}$ and $%
P(\phi _{k})=\mathbf{c}_{k}^{T}\mathbf{\Omega c}_{k}$ for a matrix $\mathbf{%
\Omega }$ that depends on $\mathbf{\beta }$ and is derived in the Technical
Supplement.

The smoothing parameters and the number of components $p$ can be chosen by
cross-validation, by maximizing 
\begin{equation}
\func{CV}(\nu _{1},\nu _{2},p)=\sum_{i=1}^{n}\log f(x_{i};\mathbf{\hat{\theta%
}}_{(-i)}),  \label{eq:cv_crit}
\end{equation}%
where $\mathbf{\hat{\theta}}_{(-i)}$ is the estimator for the reduced sample
obtained after deleting $x_{i}$.

\section{\label{sec:Example}Applications}

\subsection{Internet auction data}

In this section we analyze eBay auction data for Palm M515 Personal Digital
Assistants (PDA) on week-long auctions that took place between March and May
of 2003. The data was downloaded from the companion website of Jank and
Shmueli (2010). There were 194 auctioned items in this sample; a subsample
of 20 bid price trajectories are shown in Figure \ref{fig:data}. The dots
are the actual bids; the solid lines are for better visualization only.
Individual trajectories are hard to follow in Figure \ref{fig:data}, but
some general trends are visible. For example, bidding activity seems to
concentrate at the beginning and at the end of the auctions, in patterns
that have been called \textquotedblleft early bidding\textquotedblright\ and
\textquotedblleft bid sniping\textquotedblright , respectively. In this
paper we are interested in the bidding times as a temporal point process,
not on the bidding prices (the relationship between the two is explored in
Gervini and Baur (2017) via additive models).

\FRAME{ftbpFU}{3.7758in}{2.5452in}{0pt}{\Qcb{Internet Auction Data. Price
trajectories of Palm Digital Assistants auctioned at eBay (first 20
trajectories in a sample of 194).}}{\Qlb{fig:data}}{ebay_data.eps}{\special%
{language "Scientific Word";type "GRAPHIC";maintain-aspect-ratio
TRUE;display "USEDEF";valid_file "F";width 3.7758in;height 2.5452in;depth
0pt;original-width 7.0854in;original-height 4.7573in;cropleft "0";croptop
"1";cropright "1";cropbottom "0";filename
'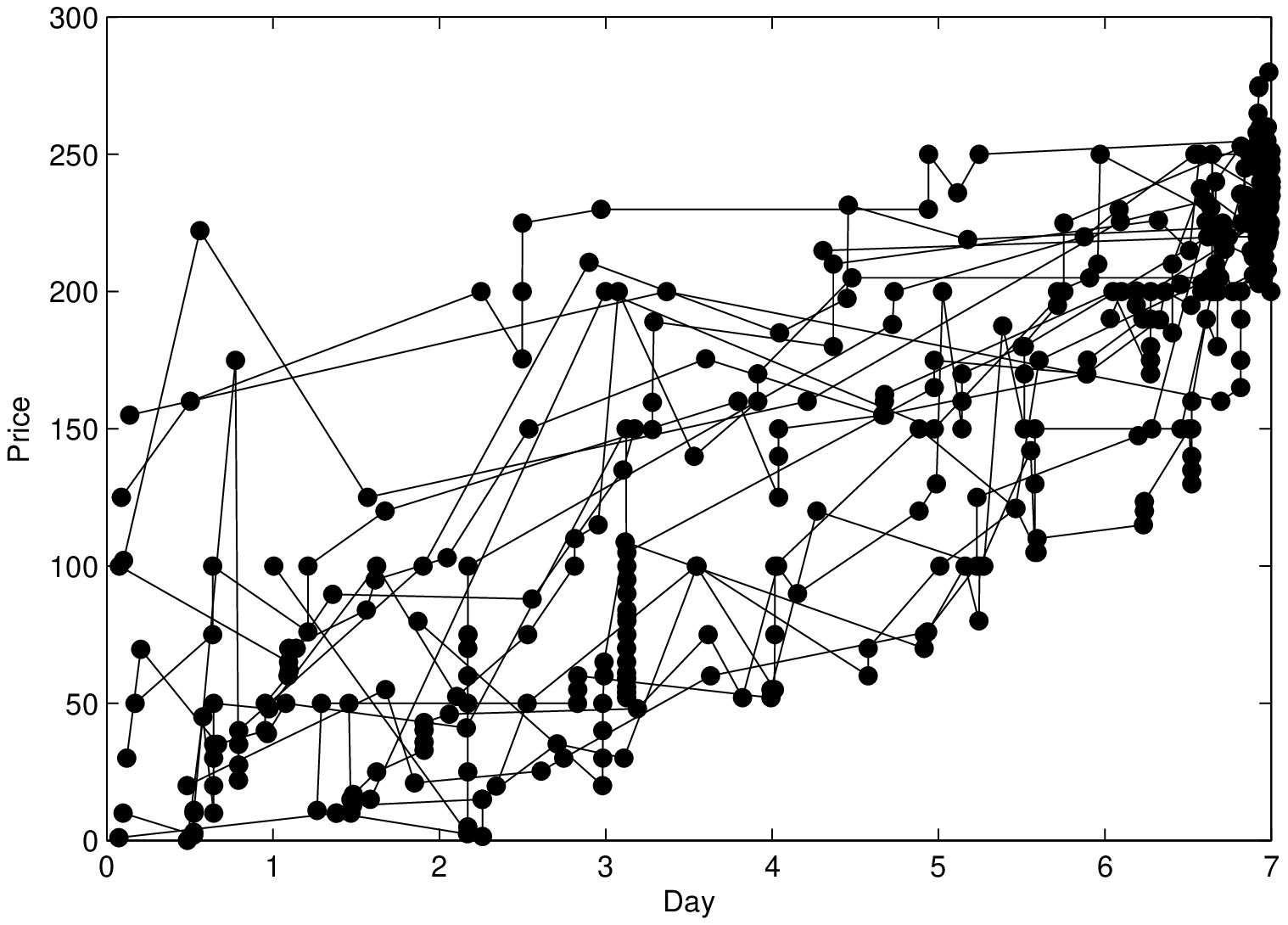';file-properties "XNPEU";}}

For these data we fitted a model (\ref{eq:Lambda_model}) with $p=2$
components, using cubic B-splines with 10 equally spaced knots as basis $%
\mathbf{\beta }$. We found the smoothing parameters $\nu _{1}$ and $\nu _{2}$
by cross-validation, obtaining $\nu _{1}=10^{-4.5}$ and $\nu _{2}=10^{-2}$.
We did not attempt to find an optimal $p$ by cross-validation, since for
illustrative purposes $p=2$ suffices. The resulting baseline intensity
function $\lambda _{0}$ and components $\xi _{1}$ and $\xi _{2}$ are shown
in Figure \ref{fig:ebay_estim}. We see in Figure \ref{fig:ebay_estim}(a)
that, as mentioned above, bidding generally intensifies towards the end of
the auction period. The component $\xi _{1}$, shown in Figure \ref%
{fig:ebay_estim}(b), is greater than one everywhere, so it is a size
component: items with component scores $u_{i1}>0$ will tend to have
intensity functions $\lambda _{i}$ that are overall larger than the baseline 
$\lambda _{0}$, so they are items that attracted lots of bidders; whereas
items with $u_{i1}<0$ will tend to have $\lambda _{i}$s overall smaller than
the baseline and therefore are items that attracted few bidders. This
interpretation is in fact corroborated by the correlation between $%
\{u_{i1}\} $ and the number of bids per item, $\{m_{i}\}$, which is $.88$.

The second component, $\xi _{2}$, is a contrast or shape component, because $%
\xi _{2}(t)>1$ for $t<1$ or $t>4$, and $\xi _{2}(t)<1$ for $1<t<4$, roughly.
So, for an item $i$ with $u_{i2}>0$, the intensity $\lambda _{i}$ will tend
to be below the baseline for $t\in (1,4)$ and above the baseline for $%
t\notin (1,4)$. In particular, items subject to strong \textquotedblleft bid
snipping\textquotedblright\ will tend to have positive $u_{i2}$s while items
that show more \textquotedblleft early bidding\textquotedblright\ will tend
to have negative $u_{i2}$s.

\FRAME{ftbpFU}{6.1393in}{2.2018in}{0pt}{\Qcb{Internet Auction Data. (a)
Baseline intensity function $\protect\lambda _{0}$. (b) Multiplicative
components $\protect\xi _{1}$ (solid line) and $\protect\xi _{2}$ (dashed
line).}}{\Qlb{fig:ebay_estim}}{ebay_lmb0_xis.eps}{\special{language
"Scientific Word";type "GRAPHIC";maintain-aspect-ratio TRUE;display
"USEDEF";valid_file "F";width 6.1393in;height 2.2018in;depth
0pt;original-width 8.8168in;original-height 2.898in;cropleft
"0.0758";croptop "1";cropright "1";cropbottom "0";filename
'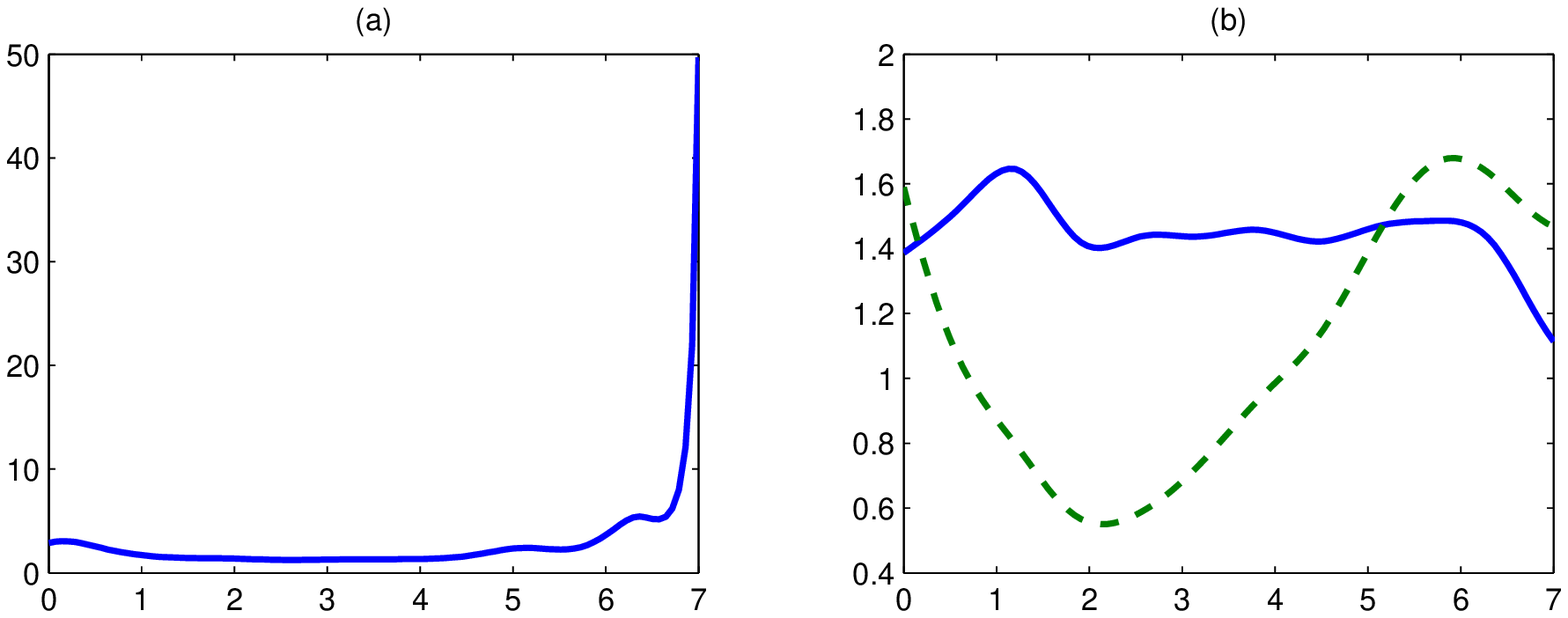';file-properties "XNPEU";}}

\subsection{Street theft in Chicago}

As a second example, this time of a spatial process, we analyzed the spatial
distribution of street robberies in Chicago during the year 2014. The data
was downloaded from the City of Chicago Data Portal, a very extensive data
repository that provides, among other things, detailed information about
every crime reported in the city. The information provided includes type,
date, time, and coordinates (latitude and longitude) of the incident. Here
we focus on crimes typified as of primary type \textquotedblleft
theft\textquotedblright\ and location \textquotedblleft
street\textquotedblright . There were 16,278 reported incidents of this type
between January 1, 2014 and December 31, 2014. Their locations cover most of
the city, as shown in Figure \ref{fig:Maps}(a); a kernel-density estimator
of these data is shown in Figure \ref{fig:Maps}(b).

\FRAME{ftbpFU}{5.9439in}{4.8948in}{0pt}{\Qcb{Chicago Street Theft. (a)
Location of reported incidents in the year 2014. (b) Kernel density
estimator of the data in (a).}}{\Qlb{fig:Maps}}{maps_color.eps}{\special%
{language "Scientific Word";type "GRAPHIC";maintain-aspect-ratio
TRUE;display "ICON";valid_file "F";width 5.9439in;height 4.8948in;depth
0pt;original-width 6.7196in;original-height 4.8672in;cropleft
"0.0596";croptop "1";cropright "0.9402";cropbottom "0";filename
'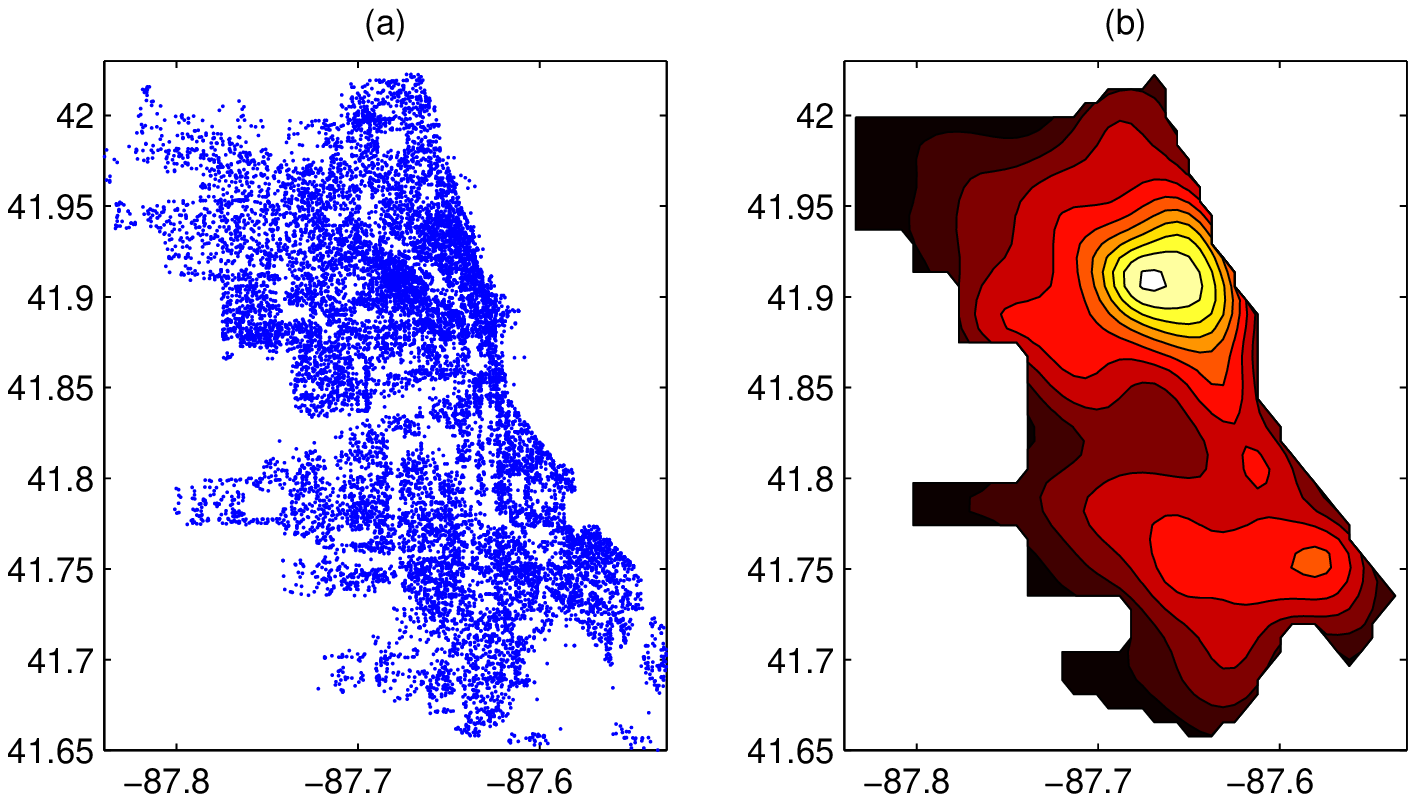';file-properties "XNPEU";}}

We grouped up the data by day and considered them as $n=365$ replications of
a spatial point process, for which we fitted a multiplicative model (\ref%
{eq:Lambda_model}). For illustrative purposes, we fitted a model with $p=3$
components (we did not attempt to find an optimal $p$). As basis $\mathbf{%
\beta }$ we used renormalized Gaussian radial kernels $\beta _{k}(\mathbf{t}%
)=\exp \{-\left\Vert \mathbf{t}-\mathbf{\tau }_{k}\right\Vert ^{2}/2\delta
_{k}^{2}\}/\sum_{j=1}^{q}\exp \{-\left\Vert \mathbf{t}-\mathbf{\tau }%
_{j}\right\Vert ^{2}/2\delta _{j}^{2}\}$, where the $\mathbf{\tau }_{k}$s
were initially 100 uniformly spaced points in $[-87.84,-87.53]\times \lbrack
41.65,42.03]$, the smallest rectangle that includes the domain $B$ (the city
of Chicago), but those $\mathbf{\tau }_{k}$s outside $B$ were eliminated,
leaving $q=40$ basis functions. The parameter $\delta _{k}$ was taken as
half the distance between $\tau _{k}$ and the closest $\tau _{j}$. The
optimal smoothing parameters were obtained by cross-validation, $\nu
_{1}=10^{-6.5}$ and $\nu _{2}=10^{-6}$.

\FRAME{ftbpFU}{6.2137in}{2.546in}{0pt}{\Qcb{Chicago Street Theft. (a)
Lateral view and (b) top view of baseline intensity function $\protect%
\lambda _{0}$. }}{\Qlb{fig:Crime_lmb0}}{crime_lmb0.eps}{\special{language
"Scientific Word";type "GRAPHIC";maintain-aspect-ratio TRUE;display
"USEDEF";valid_file "F";width 6.2137in;height 2.546in;depth
0pt;original-width 9.7222in;original-height 3.9574in;cropleft "0";croptop
"1";cropright "1";cropbottom "0";filename '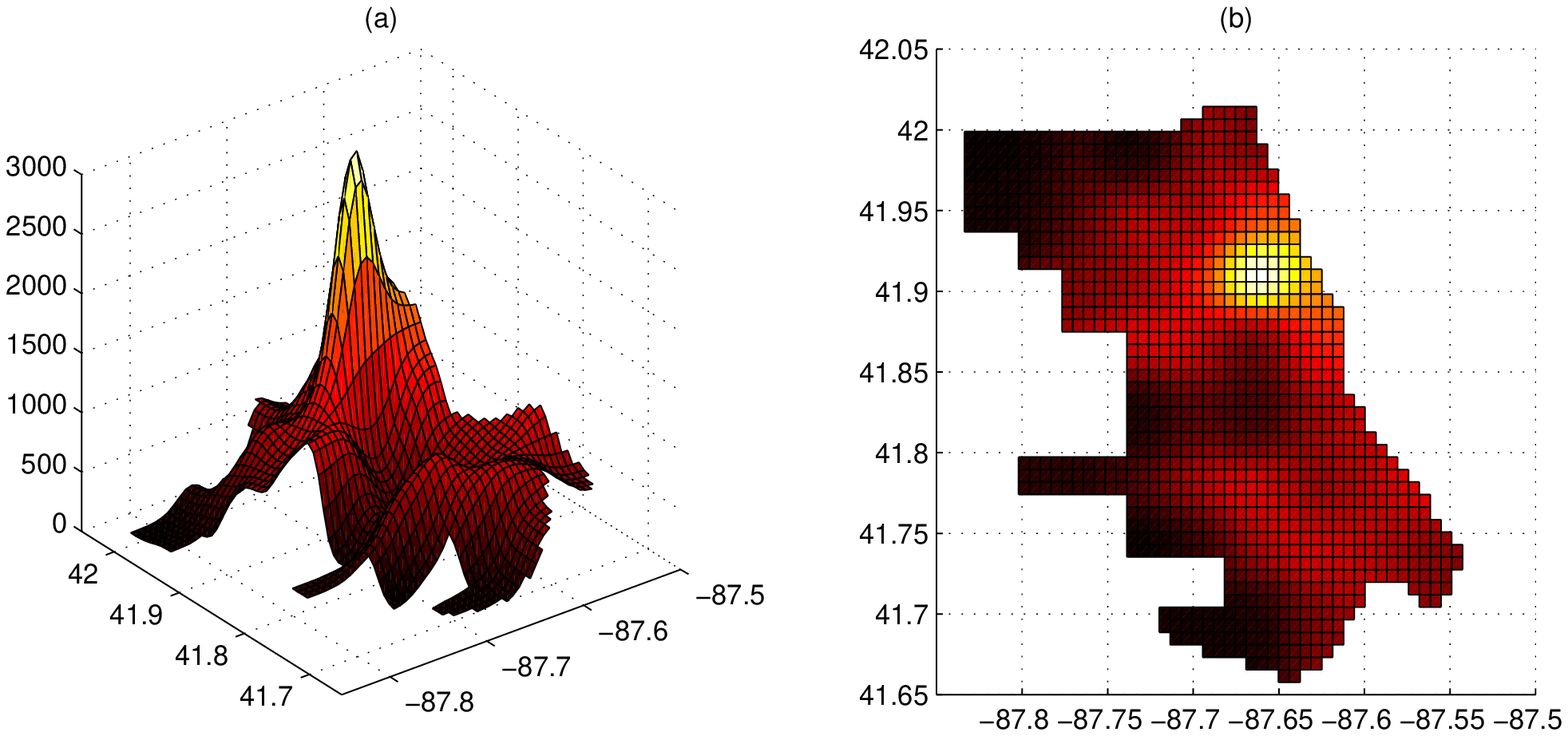';file-properties
"XNPEU";}}

The baseline intensity $\lambda _{0}$ is shown in Figure \ref{fig:Crime_lmb0}
and essentially coincides with the kernel smoother of the aggregated data
(Figure \ref{fig:Maps}(b)), as is to be expected. The mode of $\lambda _{0}$
occurs at Pulaski and Wicker Park, which are generally safe and affluent
neighborhoods, but this is precisely what attracts street thieves; the
poorer, crime-riddled neighborhoods of the West and South sides of the city
are less populated and have less foot traffic, so street theft is actually
rarer there.

The multiplicative components $\xi _{1}$, $\xi _{2}$ and $\xi _{3}$ are
shown in Figures \ref{fig:Crime_xi1}, \ref{fig:Crime_xi2} and \ref%
{fig:Crime_xi3}, respectively. The corresponding components of the
log-intensity, $\phi _{1}$, $\phi _{2}$ and $\phi _{3}$, are shown in Figure %
\ref{fig:Crime_phis}. The latter are sometimes easier to interpret due to
their scale. For instance, we clearly see that $\phi _{1}$ is nonnegative
everywhere, whereas it is not easy to determine from Figure \ref%
{fig:Crime_xi1} if $\xi _{1}$ is greater than one everywhere or not. It also
helps interpretation to plot the baseline intensity $\lambda _{0}$ versus $%
\lambda _{+}=\exp (\mu +2\sigma _{k}\phi _{k})$ and $\lambda _{-}=\exp (\mu
-2\sigma _{k}\phi _{k})$, since this shows the overall effect on $\lambda $
of moving in the direction of the components. For the first component this
is shown in Figure \ref{fig:Crime_plusmin_1}. This plot confirms that $\xi
_{1}$ is a size component: $\lambda $ will be greater than $\lambda _{0}$
everywhere for positive scores and smaller than $\lambda _{0}$ everywhere
for negative scores, and the difference in amplitude will be more noticeable
in the South-eastern part of the city, but not only in this part, as Figure %
\ref{fig:Crime_xi1} may seem to indicate. To further corroborate this
interpretation, Figure \ref{fig:Crime_days_pc1} shows the incidents in the
days with highest and lowest scores on the first component, which is in line
with what has been said.

A similar analysis reveals that the second and third components are
contrasts. For the second component, we see in Figure \ref%
{fig:Crime_plusmin_2} that positive scores correspond to $\lambda $s that
are above the baseline in the North-west part of the city and below the
baseline in the South side, and the other way around for negative scores.
The individual plots of the two extreme days (Figure \ref{fig:Crime_days_pc2}%
) confirms this. For the third component, Figure \ref{fig:Crime_plusmin_3}
shows that positive scores correspond to $\lambda $s that are above the
baseline in the narrow strip of affluent North-east neighborhoods by the
lake and below the baseline everywhere else, and the other way around for
negative scores. This is confirmed by the individual plots of the two
extreme days (Figure \ref{fig:Crime_days_pc3}).

\FRAME{ftbpFU}{5.9456in}{2.5452in}{0pt}{\Qcb{Chicago Street Theft. (a)
Lateral view and (b) top view of first multiplicative component, $\protect%
\xi _{1}$. }}{\Qlb{fig:Crime_xi1}}{crime_xi1.eps}{\special{language
"Scientific Word";type "GRAPHIC";maintain-aspect-ratio TRUE;display
"USEDEF";valid_file "F";width 5.9456in;height 2.5452in;depth
0pt;original-width 10.2774in;original-height 4.3734in;cropleft "0";croptop
"1";cropright "1";cropbottom "0";filename '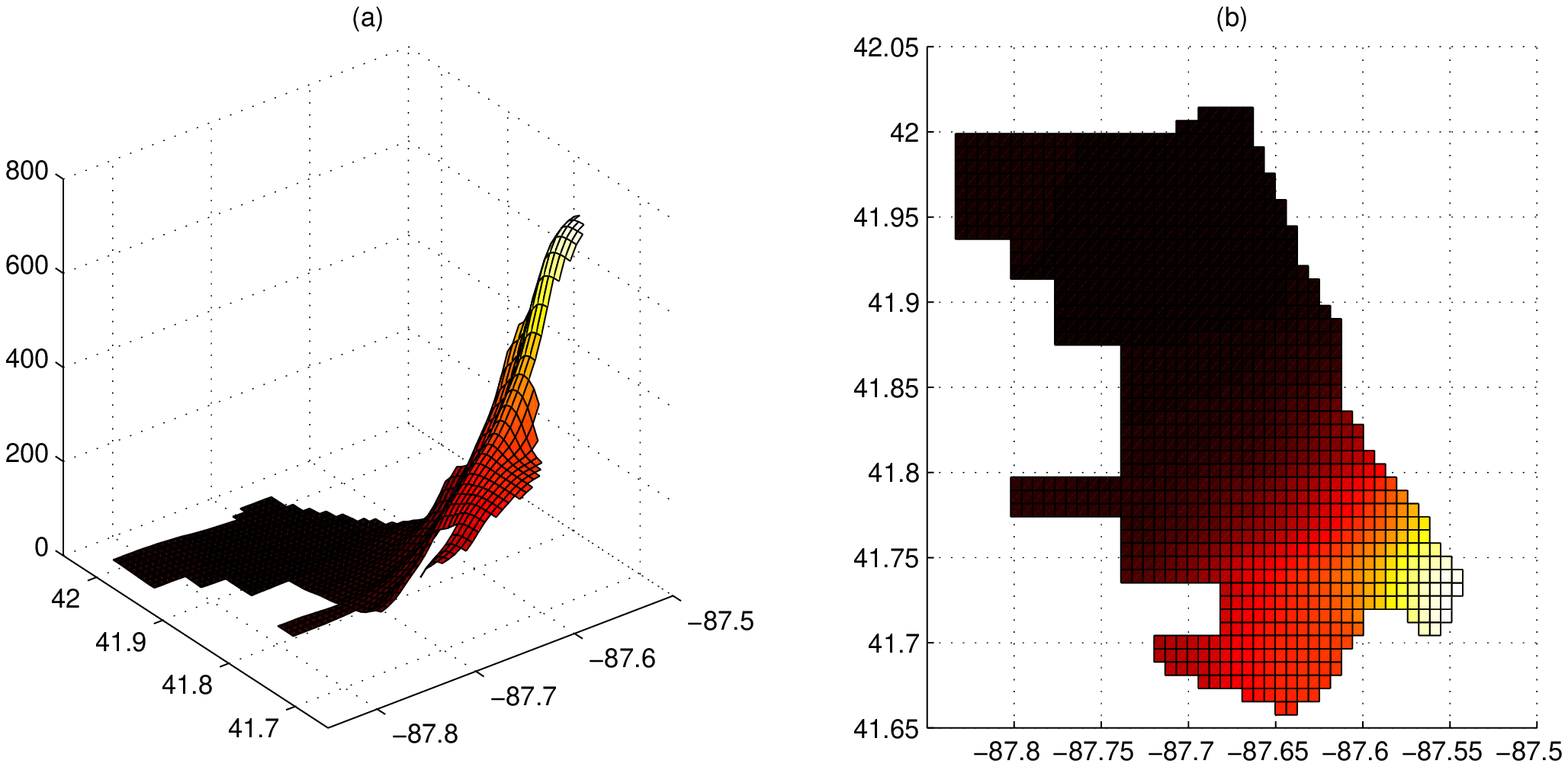';file-properties
"XNPEU";}}

\FRAME{ftbpFU}{5.9456in}{2.5452in}{0pt}{\Qcb{Chicago Street Theft. (a)
Lateral view and (b) top view of second multiplicative component, $\protect%
\xi _{2}$. }}{\Qlb{fig:Crime_xi2}}{crime_xi2.eps}{\special{language
"Scientific Word";type "GRAPHIC";maintain-aspect-ratio TRUE;display
"USEDEF";valid_file "F";width 5.9456in;height 2.5452in;depth
0pt;original-width 10.2774in;original-height 4.3734in;cropleft "0";croptop
"1";cropright "1";cropbottom "0";filename '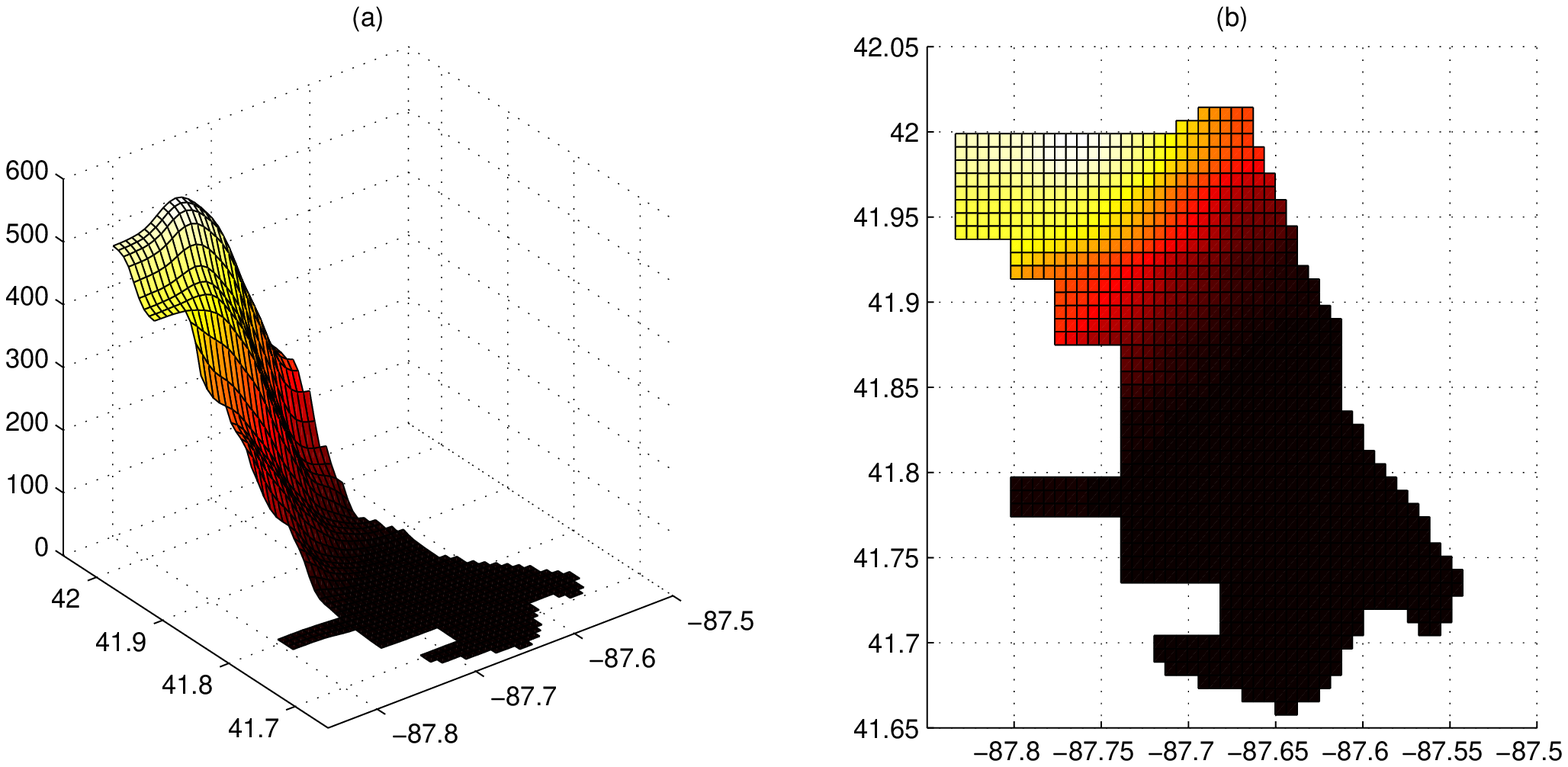';file-properties
"XNPEU";}}

\FRAME{ftbpFU}{5.9456in}{2.5452in}{0pt}{\Qcb{Chicago Street Theft. (a)
Lateral view and (b) top view of third multiplicative component, $\protect%
\xi _{3}$. }}{\Qlb{fig:Crime_xi3}}{crime_xi3.eps}{\special{language
"Scientific Word";type "GRAPHIC";maintain-aspect-ratio TRUE;display
"USEDEF";valid_file "F";width 5.9456in;height 2.5452in;depth
0pt;original-width 10.2774in;original-height 4.3734in;cropleft "0";croptop
"1";cropright "1";cropbottom "0";filename '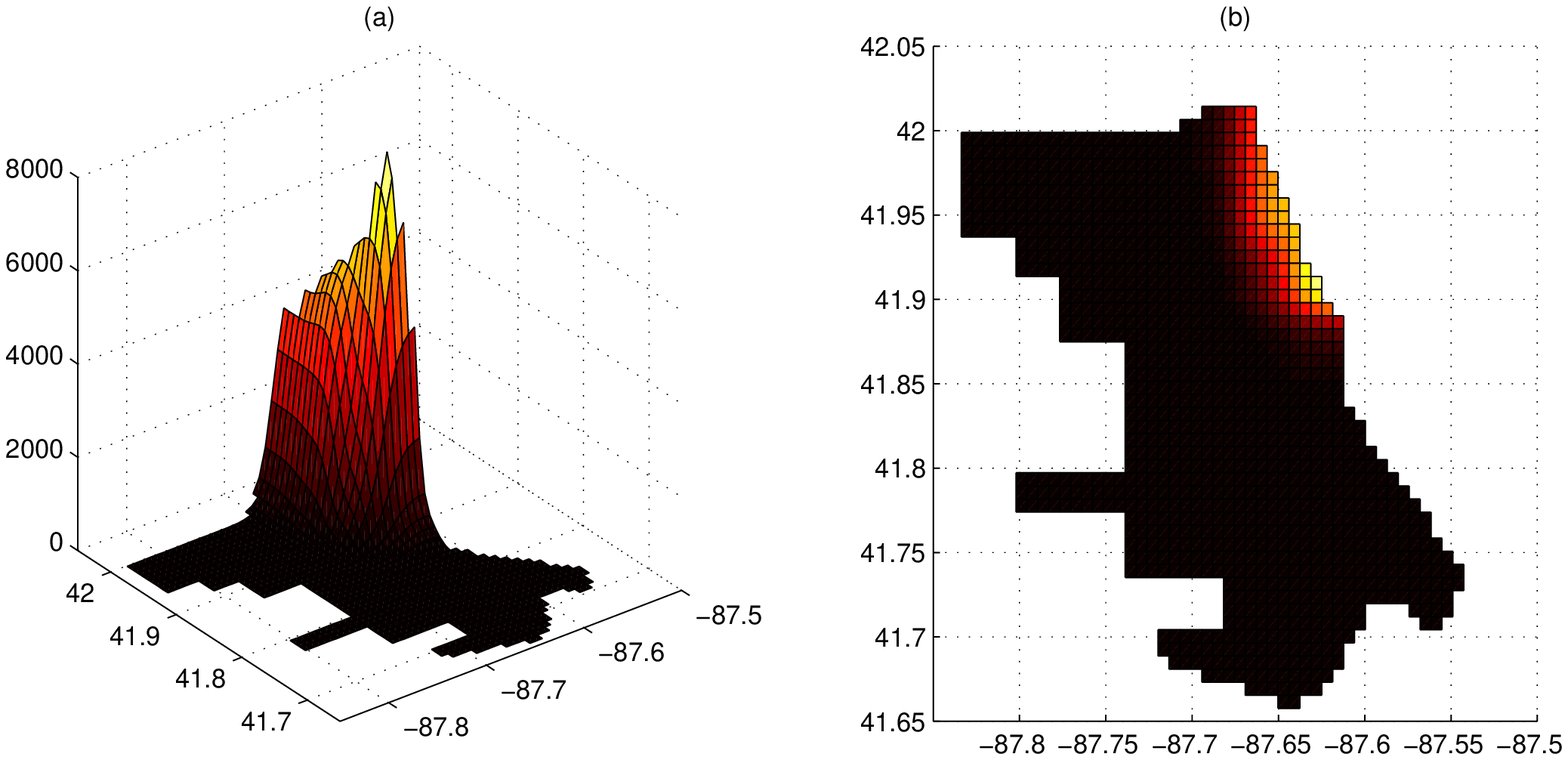';file-properties
"XNPEU";}} 

\FRAME{ftbpFU}{3.3027in}{2.5443in}{0pt}{\Qcb{ Chicago Street Theft.
Log-intensity components $\protect\phi _{1}$ (blue), $\protect\phi _{2}$
(green) and $\protect\phi _{3}$ (red).}}{\Qlb{fig:Crime_phis}}{crime_phis.eps%
}{\special{language "Scientific Word";type "GRAPHIC";maintain-aspect-ratio
TRUE;display "USEDEF";valid_file "F";width 3.3027in;height 2.5443in;depth
0pt;original-width 7.2938in;original-height 5.604in;cropleft "0";croptop
"1";cropright "1";cropbottom "0";filename '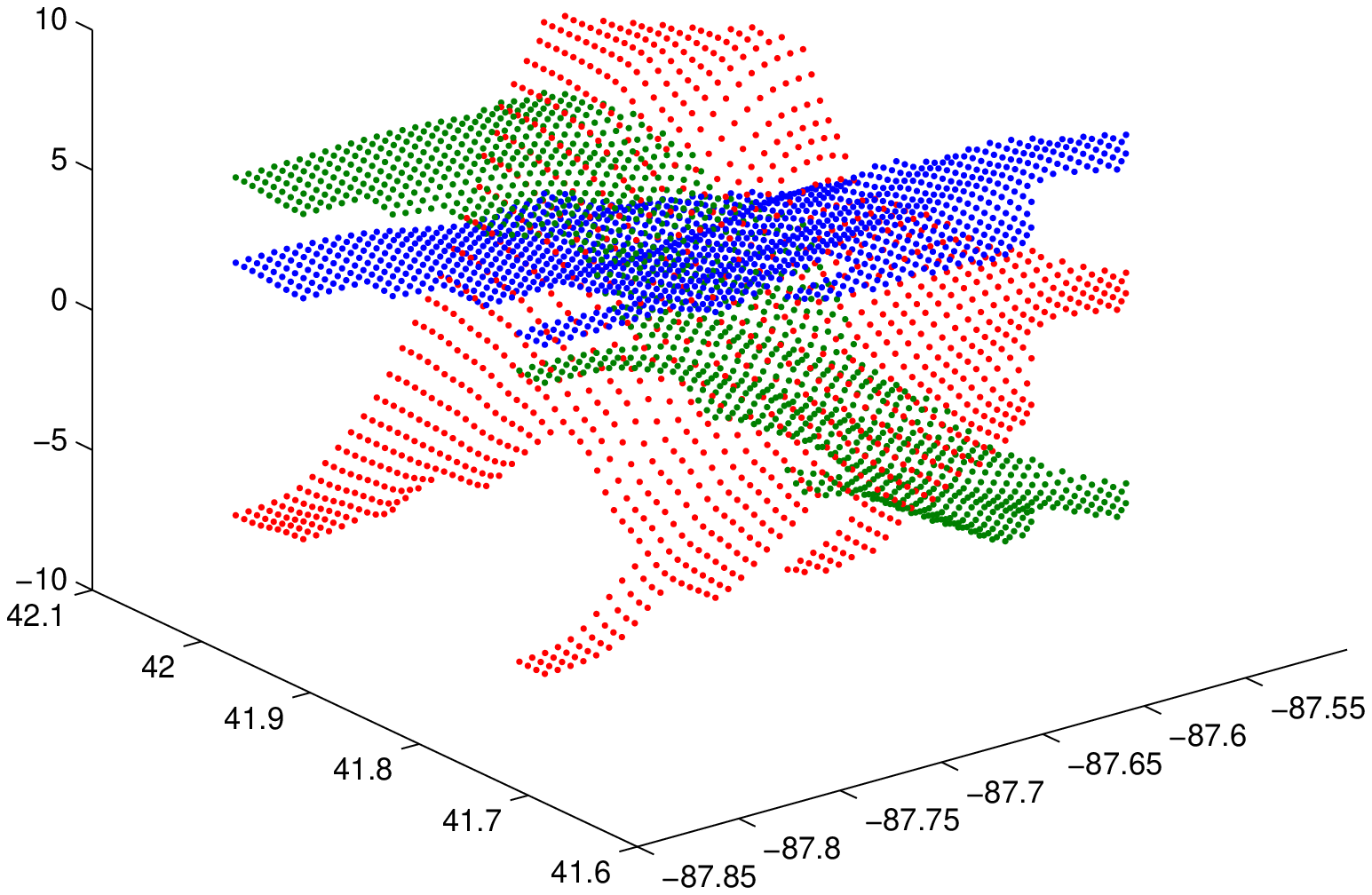';file-properties
"XNPEU";}}

\FRAME{ftbpFU}{3.0173in}{2.5443in}{0pt}{\Qcb{Chicago Street Theft. Baseline
intensity function $\protect\lambda _{0}$ (blue) versus $\protect\lambda _{-}
$ (green) and $\protect\lambda _{+}$ (red) for the first component.}}{\Qlb{%
fig:Crime_plusmin_1}}{crime_plusmin_1.eps}{\special{language "Scientific
Word";type "GRAPHIC";maintain-aspect-ratio TRUE;display "USEDEF";valid_file
"F";width 3.0173in;height 2.5443in;depth 0pt;original-width
6.6245in;original-height 5.5763in;cropleft "0";croptop "1";cropright
"1";cropbottom "0";filename '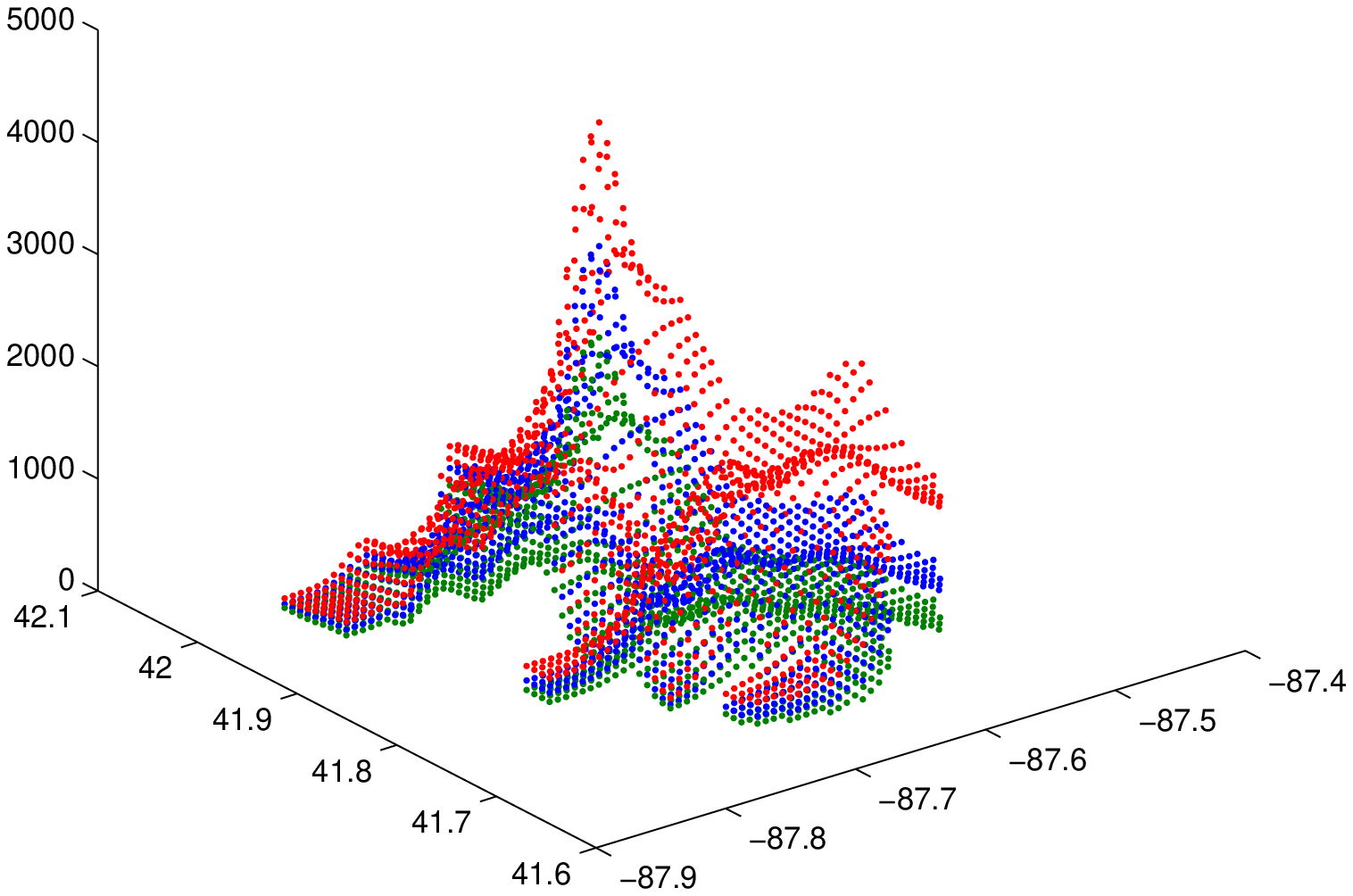';file-properties "XNPEU";}}

\FRAME{ftbpFU}{5.4379in}{2.5443in}{0pt}{\Qcb{Chicago Street Theft. Days with
highest [(a)] and lowest [(b)] scores on the first component.}}{\Qlb{%
fig:Crime_days_pc1}}{crime_days_pc1.eps}{\special{language "Scientific
Word";type "GRAPHIC";maintain-aspect-ratio TRUE;display "USEDEF";valid_file
"F";width 5.4379in;height 2.5443in;depth 0pt;original-width
9.5259in;original-height 4.4313in;cropleft "0";croptop "1";cropright
"1";cropbottom "0";filename '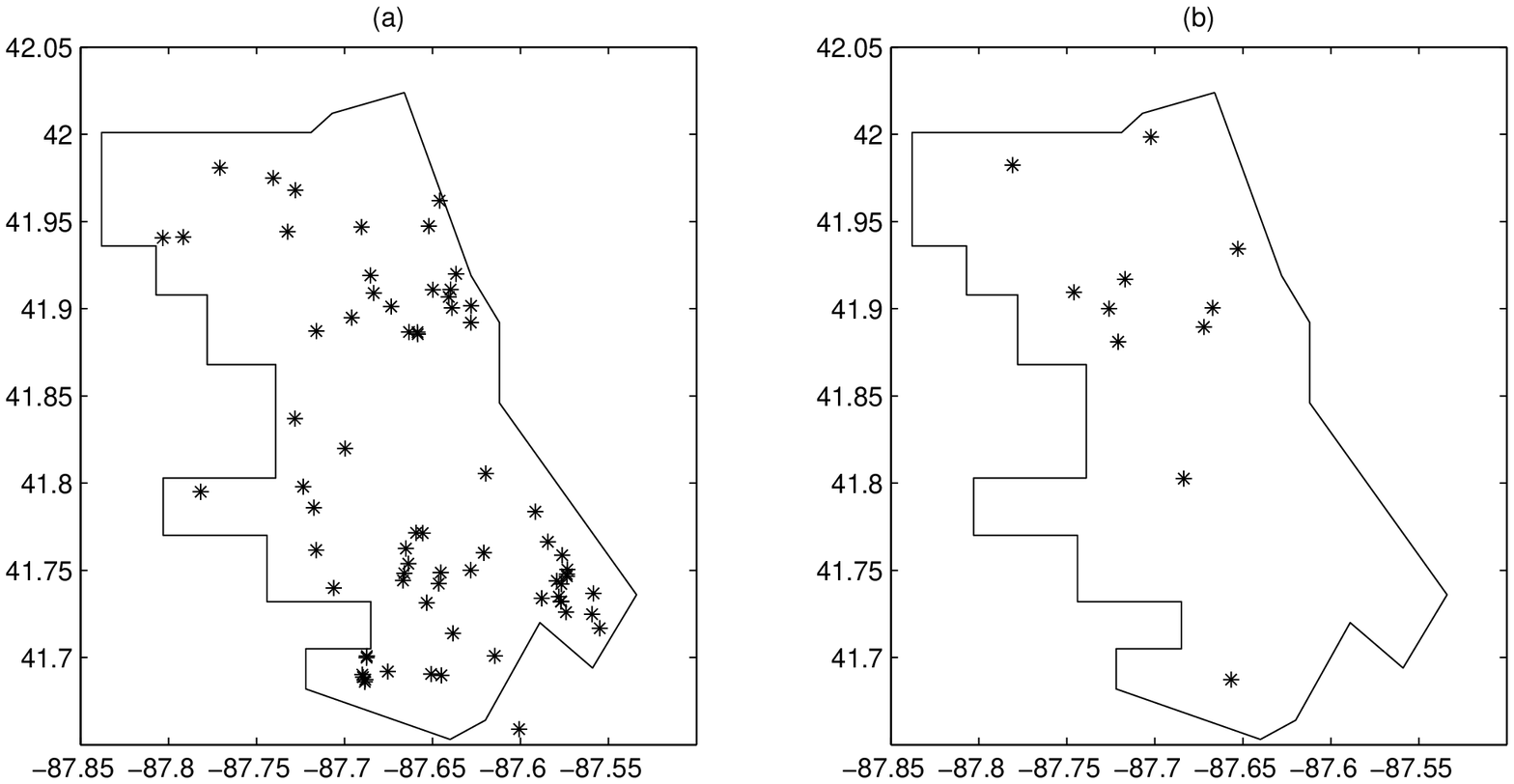';file-properties "XNPEU";}}

\FRAME{ftbpFU}{3.2759in}{2.5452in}{0pt}{\Qcb{Chicago Street Theft. Baseline
intensity function $\protect\lambda _{0}$ (blue) versus $\protect\lambda _{-}
$ (green) and $\protect\lambda _{+}$ (red) for the second component.}}{\Qlb{%
fig:Crime_plusmin_2}}{crime_plusmin_2.eps}{\special{language "Scientific
Word";type "GRAPHIC";maintain-aspect-ratio TRUE;display "USEDEF";valid_file
"F";width 3.2759in;height 2.5452in;depth 0pt;original-width
6.8753in;original-height 5.3281in;cropleft "0";croptop "1";cropright
"1";cropbottom "0";filename '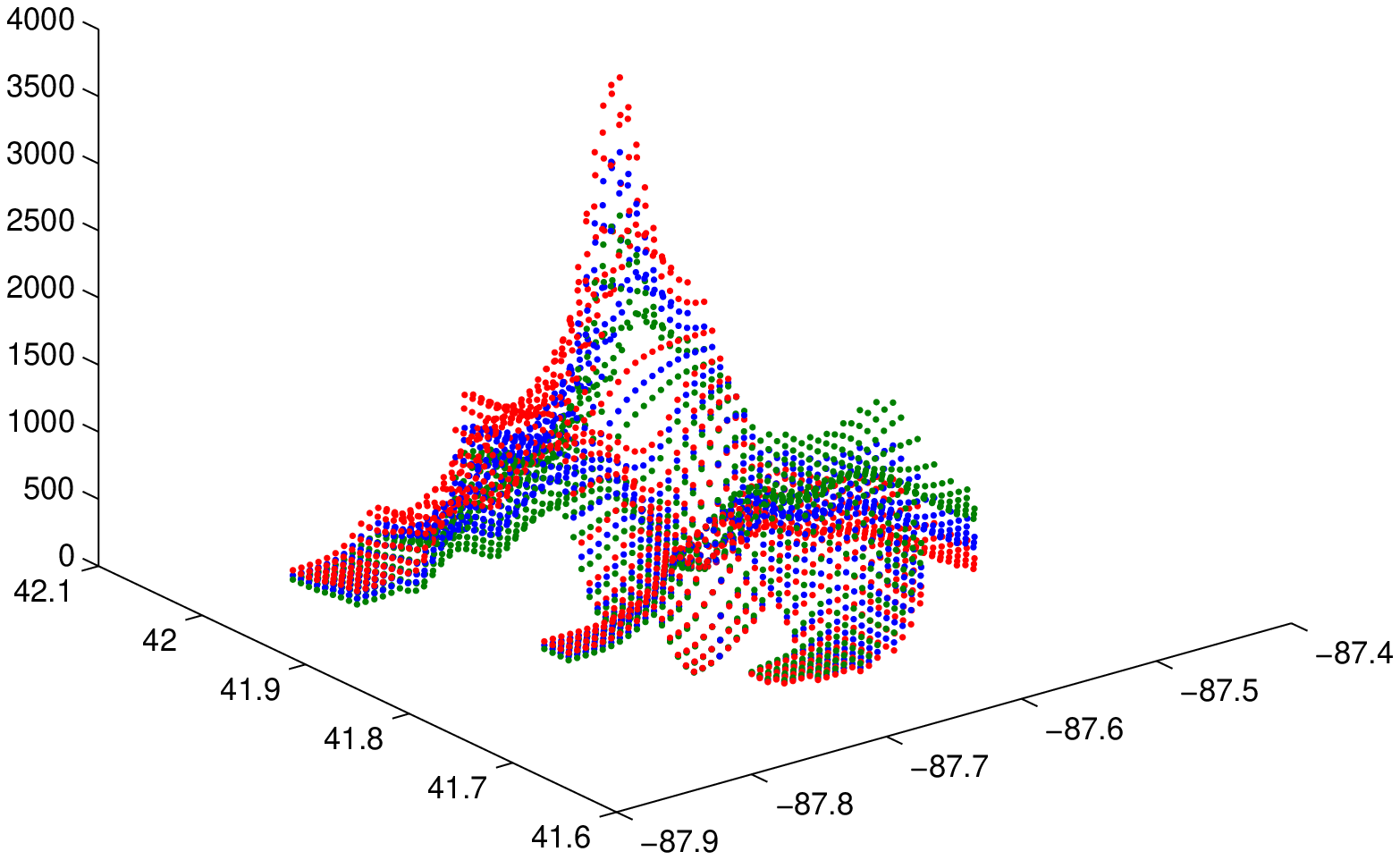';file-properties "XNPEU";}}

\FRAME{ftbpFU}{5.1612in}{2.5443in}{0pt}{\Qcb{Chicago Street Theft. Days with
highest [(a)] and lowest [(b)] scores on the second component.}}{\Qlb{%
fig:Crime_days_pc2}}{crime_days_pc2.eps}{\special{language "Scientific
Word";type "GRAPHIC";maintain-aspect-ratio TRUE;display "USEDEF";valid_file
"F";width 5.1612in;height 2.5443in;depth 0pt;original-width
9.2353in;original-height 4.529in;cropleft "0";croptop "1";cropright
"1";cropbottom "0";filename '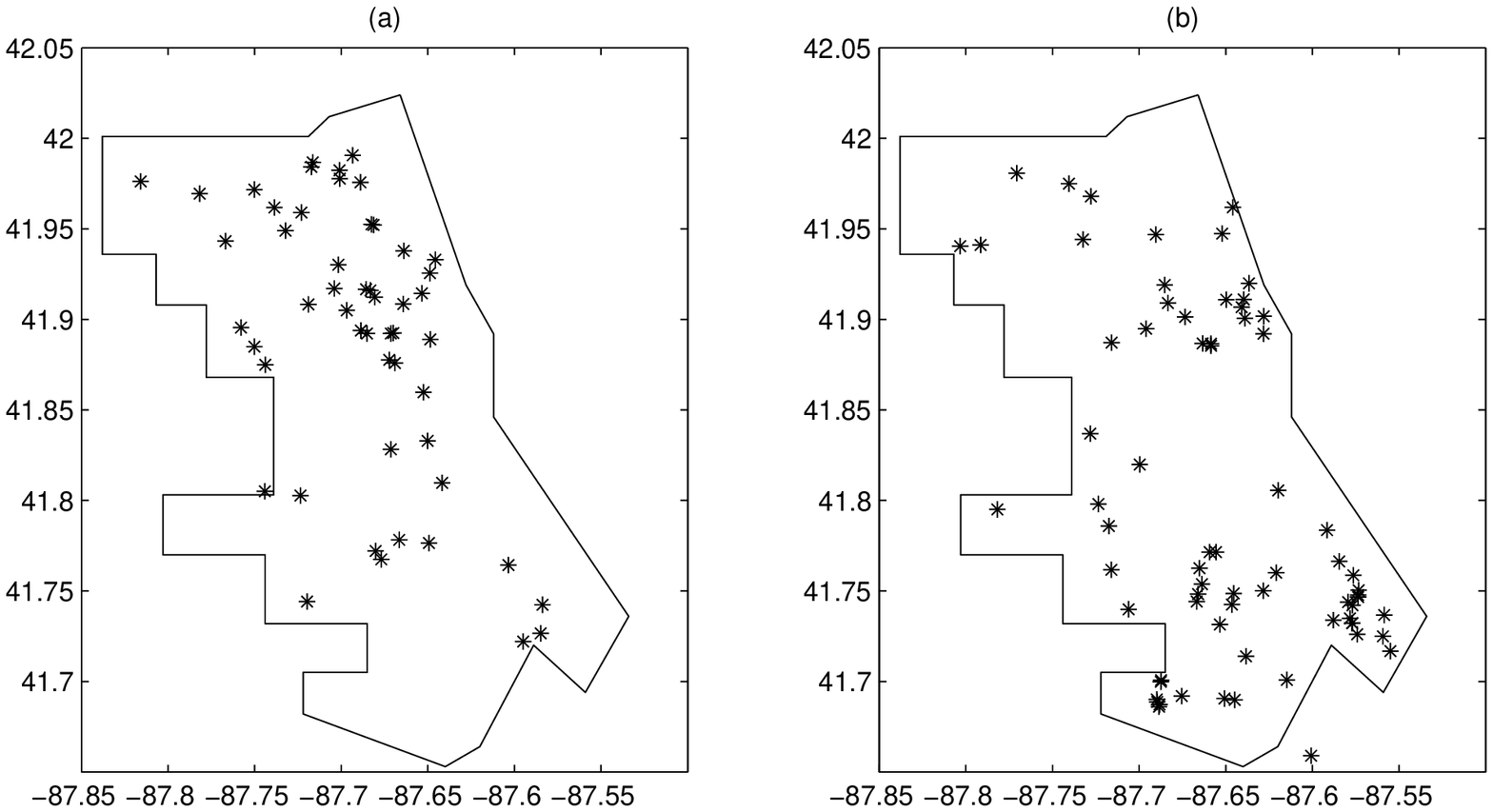';file-properties "XNPEU";}}

\FRAME{ftbpFU}{3.1946in}{2.5443in}{0pt}{\Qcb{Chicago Street Theft. Baseline
intensity function $\protect\lambda _{0}$ (blue) versus $\protect\lambda _{-}
$ (green) and $\protect\lambda _{+}$ (red) for the third component.}}{\Qlb{%
fig:Crime_plusmin_3}}{crime_plusmin_3.eps}{\special{language "Scientific
Word";type "GRAPHIC";maintain-aspect-ratio TRUE;display "USEDEF";valid_file
"F";width 3.1946in;height 2.5443in;depth 0pt;original-width
6.8597in;original-height 5.4518in;cropleft "0";croptop "1";cropright
"1";cropbottom "0";filename '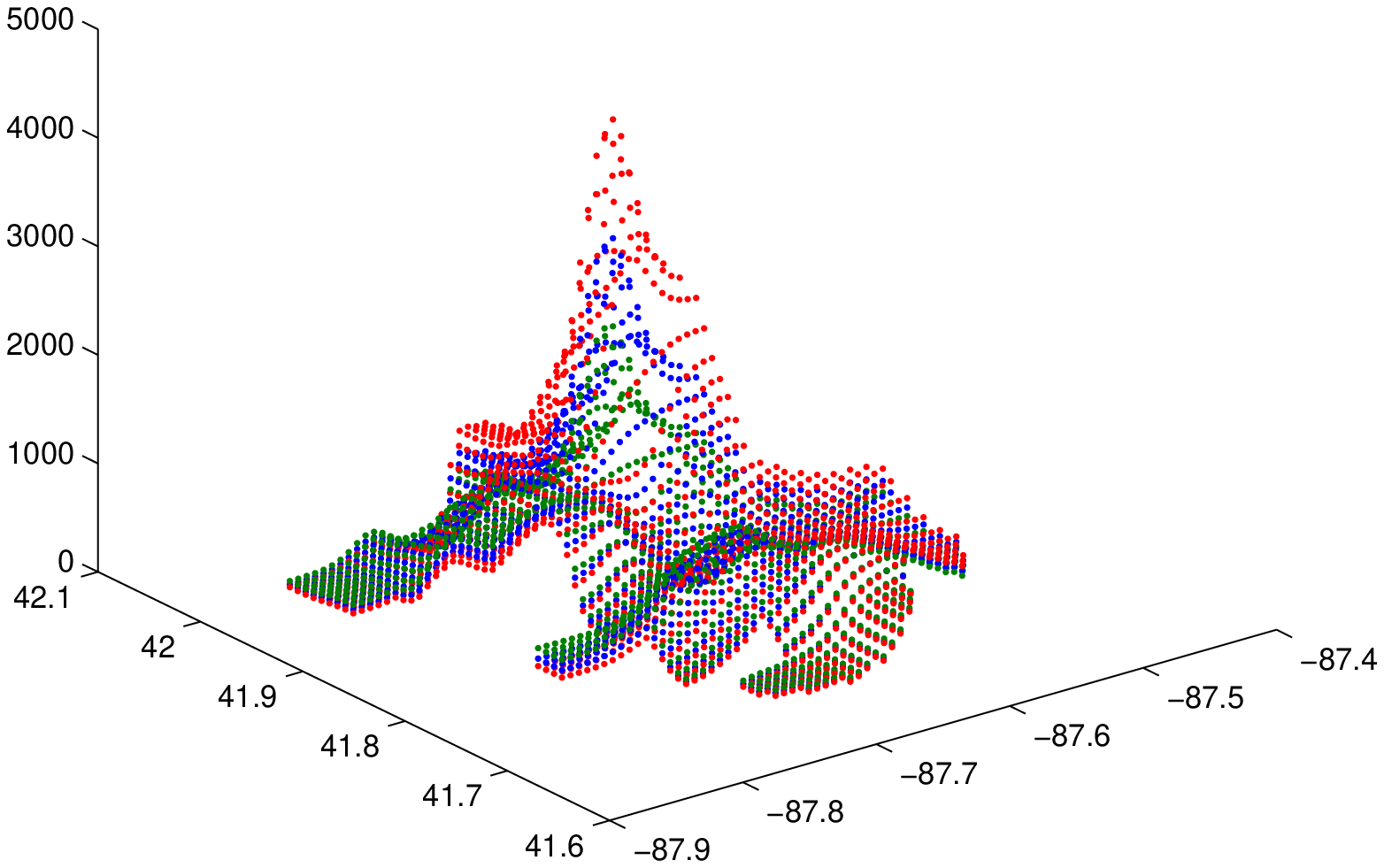';file-properties "XNPEU";}}

\FRAME{ftbpFU}{5.4665in}{2.5452in}{0pt}{\Qcb{Chicago Street Theft. Days with
highest [(a)] and lowest [(b)] scores on the third component.}}{\Qlb{%
fig:Crime_days_pc3}}{crime_days_pc3.eps}{\special{language "Scientific
Word";type "GRAPHIC";maintain-aspect-ratio TRUE;display "USEDEF";valid_file
"F";width 5.4665in;height 2.5452in;depth 0pt;original-width
9.5415in;original-height 4.4157in;cropleft "0";croptop "1";cropright
"1";cropbottom "0";filename '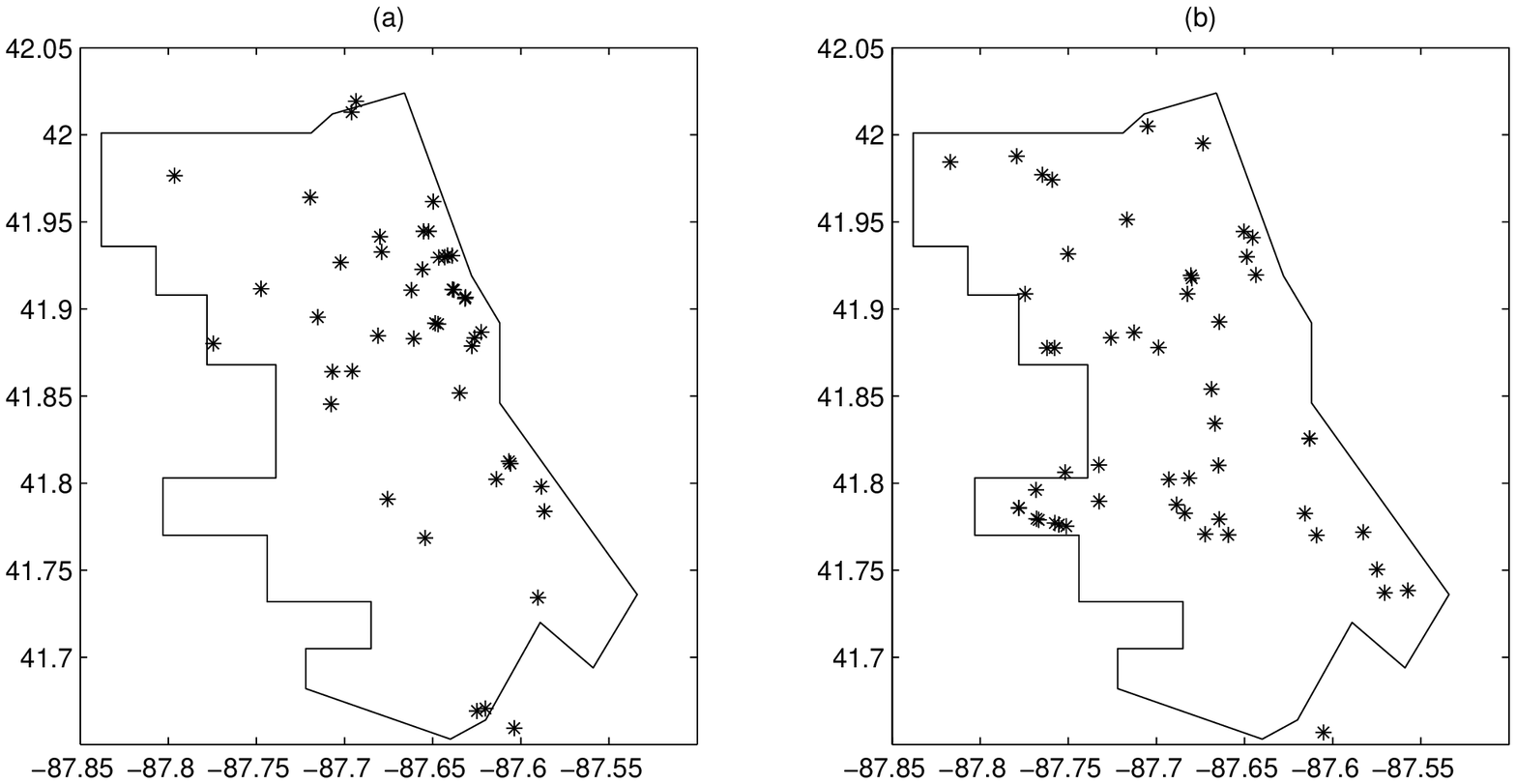';file-properties "XNPEU";}}

\section*{Acknowledgement}

This research was partly supported by US National Science Foundation grant
DMS 1505780.

\section*{References}

\begin{description}
\item Ash, R.B. and Gardner, M.F. (1975). \emph{Topics in stochastic
processes}. Academic Press, New York.

\item Baddeley, A.J., Moyeed, R.A., Howard, C.V., and Boyde, A. (1993).
Analysis of a three-dimensional point pattern with replication. \emph{%
Applied Statistics} \textbf{42 }641--668.

\item Bell, M.L., and Grunwald, G.K. (2004). Mixed models for the analysis
of replicated spatial point patterns. \emph{Biostatistics }\textbf{5 }%
633--648.

\item Cox, D.R., and Isham, V. (1980). \emph{Point Processes.} Chapman and
Hall/CRC, Boca Raton.

\item Diggle, P.J. (2013). \emph{Statistical Analysis of Spatial and
Spatio-Temporal Point Patterns, Third Edition.} Chapman and Hall/CRC, Boca
Raton.

\item Diggle, P.J., Lange, N., and Bene\v{s}, F.M. (1991). Analysis of
variance for replicated spatial point patterns in clinical neuroanatomy. 
\emph{Journal of the American Statistical Association }\textbf{86} 618--625.

\item Diggle, P.J., Mateau, J., and Clough, H.E. (2000). A comparison
between parametric and nonparametric approaches to the analysis of
replicated spatial point patterns. \emph{Advances in Applied Probability} 
\textbf{32 }331--343.

\item Diggle, P.J., Eglen, S.J., and Troy, J.B. (2006). Modeling the
bivariate spatial distribution of amacrine cells. In \emph{Case Studies in
Spatial Point Process Modeling}, eds.~A. Baddeley et al., New York:
Springer, pp.~215--233.

\item Gervini, D. (2016). Independent component models for replicated point
processes. \emph{Spatial Statistics} \textbf{18} 474-488.

\item Gervini, D. and Baur, T.J. (2017). Regression models for replicated
marked point processes. \emph{ArXiv} 1705.06259.

\item Jalilian, A., Guan, Y., and Waagpetersen, R. (2013). Decomposition of
variance for spatial Cox processes. \emph{Scandinavian Journal of Statistics 
}\textbf{40 }119--137.

\item Jank, W., and Shmueli, G. (2010). \emph{Modeling Online Auctions.}
Wiley \& Sons, New York.

\item Landau, S., Rabe-Hesketh, S., and Everall, I.P. (2004). Nonparametric
one-way analysis of variance of replicated bivariate spatial point patterns. 
\emph{Biometrical Journal }\textbf{46 }19--34.

\item Li, Y., and Guan, Y. (2014). Functional principal component analysis
of spatiotemporal point processes with applications in disease surveillance. 
\emph{Journal of the American Statistical Association }\textbf{109}
1205--1215.

\item M\o ller, J., and Waagepetersen, R.P. (2004). \emph{Statistical
Inference and Simulation for Spatial Point Processes}. Chapman and Hall/CRC,
Boca Raton.

\item Pawlas, Z. (2011). Estimation of summary characteristics from
replicated spatial point processes. \emph{Kybernetika} \textbf{47} 880--892.

\item Snyder, D.L., and Miller, M.I. (1991). \emph{Random Point Processes in
Time and Space.} Springer, New York.

\item Streit, R.L. (2010). \emph{Poisson Point Processes: Imaging, Tracking,
and Sensing.} Springer, New York.

\item Wager, C.G., Coull, B.A., and Lange, N. (2004). Modelling spatial
intensity for replicated inhomogeneous point patterns in brain imaging. 
\emph{Journal of the Royal Statistical Society Series B} \textbf{66 }%
429--446.

\item Wu, S., M\"{u}ller, H.-G., and Zhang, Z. (2013). Functional data
analysis for point processes with rare events. \emph{Statistica Sinica }%
\textbf{23} 1--23.
\end{description}

\end{document}